\documentclass[letter]{aa}                   
\usepackage{graphicx}
\usepackage{natbib,txfonts}                               
\usepackage[colorlinks=true, citecolor=blue]{hyperref} 
\usepackage{url}

\def\kms{km\,s$^{-1}$}

\begin{document}

\title{The jet of BP Tau}
\authorrunning{Dodin et al.}

   \author{A.V.~Dodin, S.A.~Potanin, M.A.~Burlak, D.V.~Cheryasov, N.P.~Ikonnikova,\\ 
           S.A.~Lamzin, B.S.~Safonov, N.I.~Shatskii, A.M.~Tatarnikov
          }

   \institute{Sternberg Astronomical Institute, Lomonosov Moscow State University,
              Universitetskij prospekt 13, 119234 Moscow, Russia\\
              \email{dodin\_nv@mail.ru}
             }

   \date{Received ...; accepted ...}

  \abstract
   {A strong global magnetic field of young low-mass stars and a high accretion rate are the necessary conditions for the formation of collimated outflows (jets) from these objects. But it is still unclear whether these conditions are also sufficient. }
   {We aim to check whether BP~Tau, an actively accreting young star with a strong magnetic field, has a jet.  }
   {We carried out narrowband [\ion{S}{ii}] 672~nm imaging and spectroscopic observations of BP~Tau and its vicinity. }
   {We find that BP~Tau is a source of a Herbig-Haro flow (assigned number HH\,1181), which includes two HH objects moving from the star  in opposite directions and a micro- (counter-) jet of $\sim 1\arcsec$ projected length. The flow is oriented along position angle $59 \pm 1 \degr.$ }
   {}


   \maketitle

\section{Introduction}
 \label{sec:intro}

Classical T Tauri stars (CTTSs) are young, low-mass stars (${\rm M} \lesssim 2{\rm M}_\odot$), the observed activity of which (spectral and photometric variability, line emission, veiling continuum, etc.) is associated with the magnetospheric accretion of protoplanetary disk matter \citep{BBB-1988,Hartmann-2016}. Disk accretion onto CTTSs is accompanied by mass outflow from the disk. Jets and winds play an important role in removing angular momentum from CTTSs and their disks, thus setting the initial conditions for planet formation \citep{Frank-2014}.
The Balmer \ion{H}{i} and forbidden lines (FLs) in the spectra of CTTSs are our main source of information about the kinematics and physical conditions of the outflow. 

The profiles of FLs consist of (at least) two components: a low velocity component (LVC), $|V_{\text r}| \lesssim 30$~{\kms}, and a high velocity component (HVC), which trace spatially and physically distinct regions \citep{Kwan-Tademaru-1988,HEG-1995}. The LVC of FLs is formed in the wind {\lq\lq}blowing{\rq'} from the surface of the accretion disk due to the photoevaporation of the disk atmosphere, which is heated by UV and X-ray radiation from the central star and/or a magneto-centrifugal mechanism  (see, e.g., \citealt{Weber-2020} and references therein). The FLs of most, if not all, CTTSs have LVCs \citep{Nisini-2018}, the majority of which are stable on a timescale of decades \citep{Simon-2016}. 

The HVCs of FLs originate in the jets, that is, in extended (up to 3 pc) fast $(V \sim 300$~\kms) highly collimated bipolar gas flows \citep{Bally-2016}. The jets look like a chain of compact emission nebulae (knots), the so-called Herbig-Haro (HH) objects, with less dense gas between them. The FLs in the spectrum of a star appear to have a HVC if such a knot (micro-jet) falls into the slit of the spectrograph. Otherwise, the jet can be detected in images of the star's vicinity obtained in a narrowband filter centered at a suitable emission line, such as [\ion{S}{ii}] 6731~\AA{} or H$\alpha$.
 
There is a general consensus that jet launching and collimating involve the dynamical interaction of the innermost regions of the disk with the global magnetic field of the CTTS \citep{Ferreira-2006, Beskin-2023}. In other words, a strong global magnetic field and a high accretion rate are the necessary conditions for the formation of jets from CTTSs. But it is unclear if these conditions are also sufficient. This is a nontrivial question because 
the frequency of the jet phenomenon among CTTSs is $<40$\,\% \citep{Nisini-2018}.

BP~Tau is of particular interest in this context. This single \citep{Kounkel-2019} CTTS \citep{KH-1995} has a strong global magnetic field \citep{Donati-2008,Flores-2019} and actively accretes matter \citep{Long-2011} from its protoplanetary disk, as discovered by \citet{Long-2019}. The star is also the source of an outflow observed in emissions from forbidden \citep{HEG-1995} and dipole-allowed \citep{Gullbring-1996, Alencar-2000, Errico-2001} transitions. 

The supposed evidence of HVCs in the FLs of BP~Tau is controversial. The [\ion{S}{ii}]~6731~\AA{} line redshifted by $\approx 120$~{\kms} relative to the star was observed by \citet{HEG-1995} to have equivalent widths (EWs) of $0.10\pm 0.05$ and $0.07 \pm 0.02$~\AA{} on January 5 and 7, 1988, respectively. However, according to \citet{Simon-2016}, the EW of this line was $<0.005$~\AA{} in the spectra observed in 2006 and 2012. The [\ion{S}{ii}]~6731~\AA{} line is also absent in the high-resolution spectra of BP~Tau observed in 2021-2022 by \citet[see their Fig. A.1]{Nisini-2023}, but the authors found weak HVCs in the [\ion{O}{i}]~5577, 6300~\AA,{} and [\ion{S}{ii}]~4068~\AA{} lines, which, in their opinion, indicated the presence of a jet from BP~Tau. 

Considering all this, the task of detecting a jet from BP~Tau is worthwhile.


\section{Observations}
 \label{sec:obs}

The observations of BP~Tau were carried out with the 2.5~m telescope at the Caucasian Mountain Observatory (CMO) of the Sternberg Astronomical Institute of Lomonosov Moscow State University \citep[SAI MSU;][]{Shatsky-2020}. Most of the spectroscopic data were obtained with the Transient Double-beam Spectrograph (TDS; see \citealt{Potanin-2020} for a description of the instrument and data reduction procedures). The spectral resolution of the TDS is $R=\lambda/\Delta \lambda \approx 2400$ in the red channel $(0.56 - 0.74$~$\mu$m) and $\approx 1300$ in the blue $(0.36 - 0.56$~$\mu$m). The log of observations is given in Table~\ref{tab:log-TDS}. To obtain the absolute flux for the discovered HH objects, we also performed spectrophotometric observations under suitable conditions with the spectrophotometric standard G191-B2B.

 \begin{table}
\caption{Log of TDS observations.}             
\label{tab:log-TDS}      
\begin{tabular}{l c c r c}        
\hline\hline                 
rJD$^a$ &  Object & Exposure, s &PA, \degr & slit width, \arcsec  \\    
\hline                        
298.3504 & BP Tau & 1200& 62     & 10  \\
298.3792 & BP Tau & 1200& 17     & 10  \\
298.3960 & BP Tau & 1200& 107    & 10  \\
299.3048 & BP Tau & 2400& 42     & 10  \\
299.3357 & BP Tau & 2400& 82     & 10  \\
299.3773$^b$ & BP Tau & 4200& 62     & 10  \\
299.5599 & knot A & 2400& 18    & 1.5  \\
311.3346 & knot B & 4800& $-28$ & 1.5  \\
368.2269$^b$ & knot A & 3600& 18    & 10  \\
\hline                                   
\end{tabular}\\
(a) rJD$=$HJD$-2\,460\,000$; (b) spectrophotometric observations.
\end{table}

When we observed BP~Tau, the light input unit and the fiber line for the future high-resolution spectrograph of the 2.5~m telescope were undergoing tests with the help of an old SAI spectrograph, RADUGA (its main R2 echelle grating is 75~lines/mm, with a double-pass prism cross-dispersion, a beam of 45~mm, and a spectral range of 4100 -- 8200~\AA). The 100~$\mu$m fiber output (the equivalent of a 2\farcs5 diaphragm) was reimaged onto the slit to cut some 50\% of the light and push the resolving power up to $R\sim 24\,000$. We used this occasion to obtain spectra of BP~Tau, on January 11, 2024. 

Direct $10\arcmin \times 10\arcmin$ images of the region around BP~Tau were obtained with the NBI CCD camera of the 2.5 m telescope in two filters -- [\ion{S}{ii}]~$672$~nm, with a total exposure of 170~min, and the nearby continuum [\ion{S}{ii}]rc (115~min) -- on December~20, 2023 (JD=2\,460\,299.5) and January~4, 2024 (JD=2\,460\,314.3).\footnote{\url{https://obs.sai.msu.ru/cmo/sai25/wfi/}} To minimize the effects of variable seeing, we changed the [\ion{S}{ii}] and [\ion{S}{ii}]rc filters sequentially with individual exposures of $5-15$ min.


\section{Results and discussion}
 \label{sec:results}

\subsection{The micro-jet}
 \label{subsec:mjet}
  
   \begin{figure}
   \centering
   \includegraphics[width=\hsize]{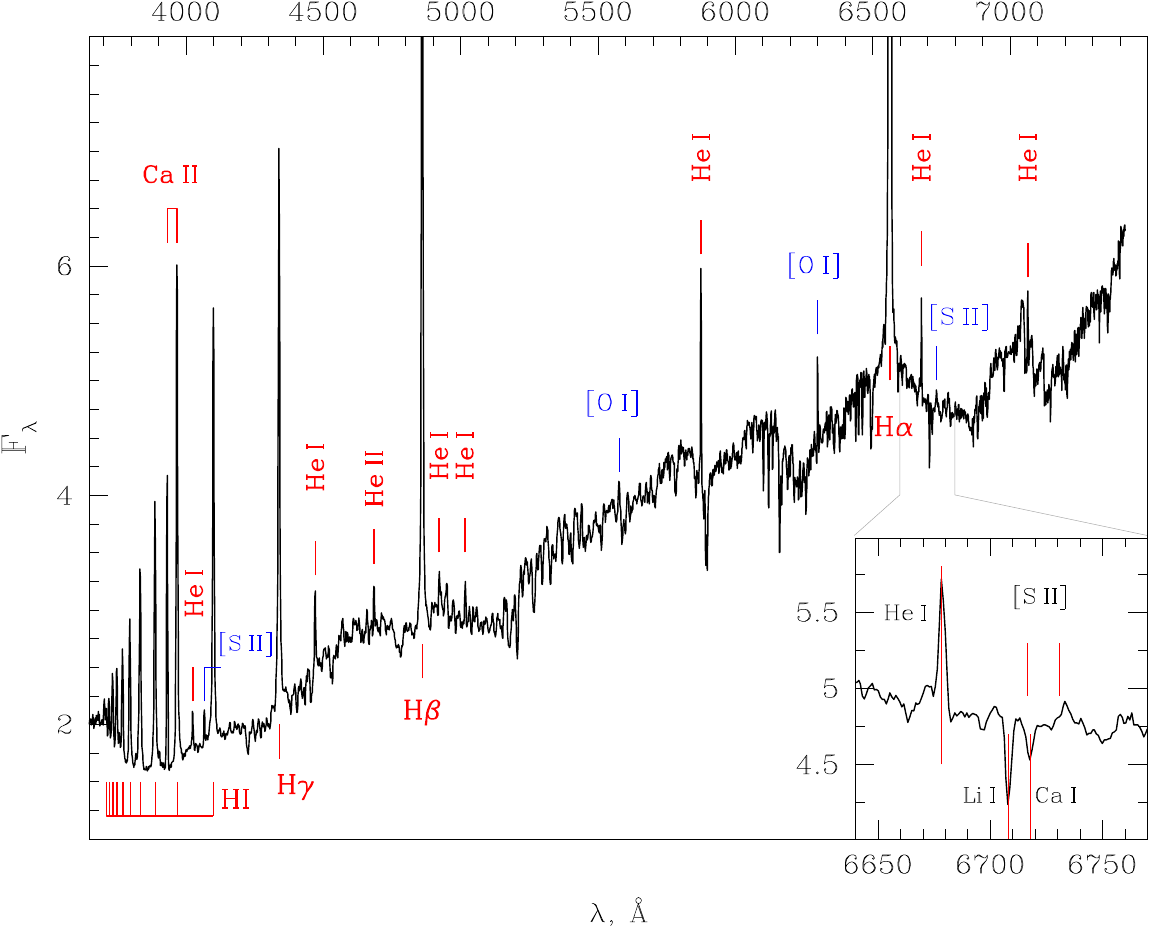}
      \caption{TDS spectrum of BP~Tau. The portion of the spectrum in the vicinity of the
      [\ion{S}{ii}]~6716, 6731~\AA{} lines is shown on a larger scale. $F_\lambda$ is the relative flux. 
              }
         \label{fig:TDS-spectrum}
   \end{figure}

The low-resolution TDS spectrum of BP~Tau observed on $\text{JD}=2\,460\,298.3504$ is shown in Fig.~\ref{fig:TDS-spectrum}. Along with strong emission lines of \ion{H}{i}, \ion{He}{i}, \ion{and Ca}{ii,} the FLs [\ion{O}{i}]~5577, 6300~\AA{} and [\ion{S}{ii}]~4068 are clearly seen. The redshifted [\ion{S}{ii}]~6731~\AA{} emission is barely noticeable, and the [\ion{S}{ii}]~6716~\AA{} emission is probably blended by the \ion{Ca}{i}~6717.7~\AA{} absorption.

\begin{figure}
   \centering
   \includegraphics[width=\hsize]{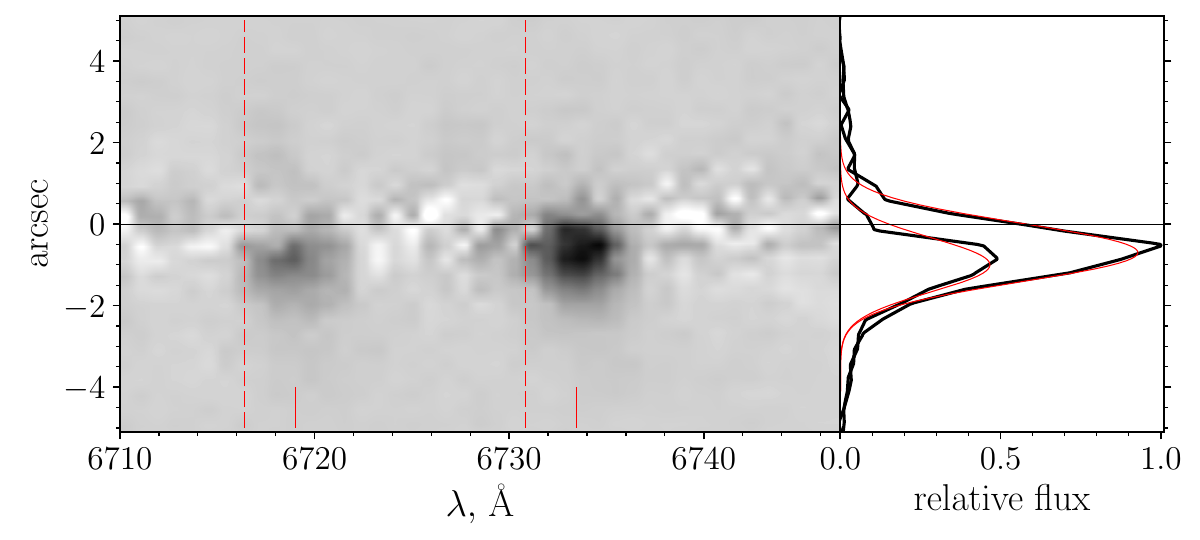}
      \caption{TDS spectrum of the micro-jet. Left panel: Long-slit TDS spectrum of BP~Tau near [\ion{S}{ii}] lines observed at ${\text PA}=62\degr$. The image is centered on the star in spatial and wavelength coordinates. The spatial profile of the stellar spectrum has been removed. Emission knots are shifted to the southwest from the star. The rest wavelengths of the [\ion{S}{ii}] lines are shown with dashed lines. The measured velocity of +118\,{\kms} is marked with solid red lines. Right panel: Relative flux distribution in the [\ion{S}{ii}]~6716 and 6731~\AA{} lines along the slit and the corresponding fits with Gaussian profiles. }
         \label{fig:cknot}
   \end{figure}

To locate the source of the [\ion{S}{ii}] emission, we obtained a TDS spectrum with a 10\arcsec-wide slit (JD=2\,460\,299.377; see Table\,\ref{tab:log-TDS}) with an expected position angle (PA) of the jet of $62\degr$, that is, nearly perpendicular to the major axis of the BP~Tau ALMA disk image \citep{Long-2019}. From the obtained 2D spectrum, we removed the spatial profile of the star reconstructed using the regions outside the emissions (see Fig.~\ref{fig:cknot}). In each line we see a knot at a distance of $\approx 0\farcs8$ from the star and at a radial velocity of $V_\text{r}^{\text mj}=118\pm 5$~{\kms} in the stellar rest frame. We integrated this 2D spectrum over the wavelength within each line of the doublet, and the corresponding spatial profiles are shown in the right panel of Fig.\,\ref{fig:cknot}. The formal approximation of these profiles gives distances of $1\farcs0$ and $0\farcs7$ for the centers of [\ion{S}{ii}]\,6716 and 6731 {\AA}, respectively. We believe that this result indicates the presence of a micro- (counter-) jet from the star. To estimate its orientation (PA, $\theta_0$) and projected length ($l$), we obtained spectra of the star at different orientations ($\theta$) of the spectrograph slit. 

\begin{figure}
   \centering
   \includegraphics[width=\hsize]{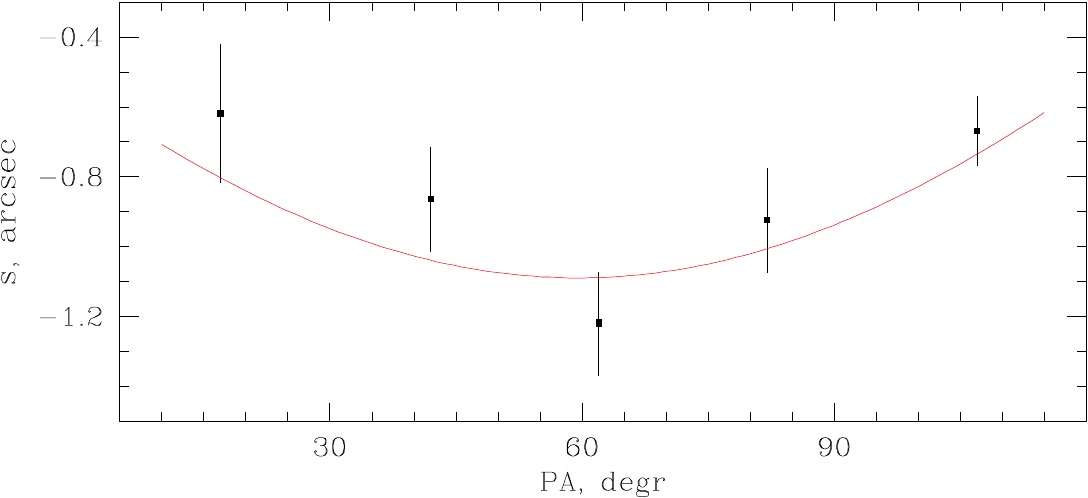}
      \caption{Observed dependence, $s=s\left( \theta \right)$, and its fit (red curve). See the main text for details. }
         \label{fig:spastrom}
   \end{figure}

If we consider the star and the micro-jet as two point sources, then the projection distance, $s,$ between them along the spectrograph slit is $s=l\cos \left[ \pi- \left( \theta_0-\theta \right) \right].$ The observed dependence, $s=s\left( \theta \right),$ is shown in Fig.~\ref{fig:spastrom} together with the fitting curve derived via the least squares method with $\sigma_s^{-2}$ as the weight. This fit yielded $\theta_0= 59\fdg5 \pm 6\fdg5,$ $l= 1\farcs09 \pm 0\farcs09.$ 

We also observed a [\ion{O}{i}] 6300~\AA{} line photocenter shift of $-9\pm 3$~mas at $V_{\text r}=+115 \pm 20$~{\kms} relative to the nearby spectrum in our TDS observation with the best signal-to-noise ratio (S/N) -- JD=2\,460\,299.377, PA=62\degr. Unfortunately, we cannot convert it to the projection distance, $s,$ due to the dominating contribution of the wind component in the line profile, so the two-component (star--jet) model cannot be applied. If we assume $s=-1\arcsec$ for the [\ion{O}{i}] line, then the amplitude of the micro-jet component at the TDS resolution is 0.01 in continuum units, or about 10\% of the wind component. We did not find any spectro-astrometric shift in the blue wing of the [\ion{O}{i}] 6300~\AA{} line.

An independent estimate of the micro-jet velocity can be obtained with higher-resolution spectra of BP~Tau. The latest available archival\footnote{\url{https://archive.eso.org}} high-resolution spectrum of the star was observed with the ESO VLT ESPRESSO spectrograph \citep{Gonzales-2018} in December 2022, but the [\ion{S}{ii}] lines are just barely detected in the spectrum, probably due to ESPRESSO's small aperture diameter, $d=1\arcsec$ , which does not cover the micro-jet located at a $l \sim 1\arcsec$ distance from the star. For the same reason, the sulfur lines are not visible in the spectra analyzed by \citet{Nisini-2023}, who also used a $d=1\arcsec$ aperture. But in our RADUGA spectrum obtained with a 2\farcs5 aperture, these lines are present, albeit very weak.

\begin{figure}
   \centering
   \includegraphics[width=\hsize]{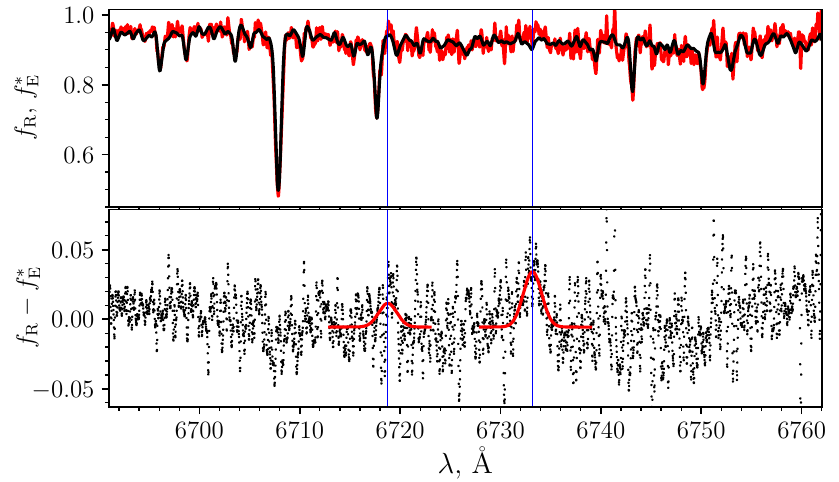}
      \caption{High-resolution spectra of BP Tau. Upper panel: RADUGA spectrum (red) and the corrected ESPRESSO spectrum (black). Lower panel: Their difference and the fit for the [\ion{S}{ii}] lines. See the main text for details.}
         \label{fig:raduga}
\end{figure}

To highlight the difference between the small-aperture ESPRESSO spectrum and our spectrum, we reduced the ESPRESSO resolution to the RADUGA resolution $(R \approx 24\,000)$ and veiled the ESPRESSO spectrum, $f=f(\lambda),$ by $r=0.3$ so that $f^*=(f+r)/(1+r),$ to adjust the depth of the absorption lines.\footnote{
\citet{Dodin-2013} demonstrate that the continuum, rather than the line emission, contributes most to the veiling of BP~Tau, so it is reasonable to assume that $r$ is almost constant within the $\sim 100$~\AA{} spectral range in the vicinity of the sulfur lines.}
The two spectra and their difference are plotted in Fig.\,\ref{fig:raduga}. We fitted the residual profiles of the sulfur lines with two Gaussians, assuming a fixed distance between them, identical widths, and a flux ratio corresponding to the high density limit, $F_{6716}/F_{6731} \approx 0.43$ \citep{Giannini-2015}, which is close to the value we found from low-resolution TDS spectra (see, e.g., the right panel of Fig.\,\ref{fig:cknot}). The final result is that the [\ion{S}{ii}] lines are redshifted by $V_{\text r}^{\text mj}=106\pm 3$\,{\kms} relative to the nearby stellar absorption lines.\ This is in reasonable agreement with the TDS measurement $V_{\text r}^{\text mj}=118\pm 5$\,{\kms}, keeping in mind the distance--velocity interplay due to the wide slit and the uncertainty in the PA of the jet.

Considering the above, it is not clear what physical connection exists between the micro-jet and the HVC features of the [\ion{S}{ii}]~4068, [\ion{O}{i}]~5577, and 6300~\AA{} lines, which trace higher critical densities \citep{Nisini-2023}.

We next estimated the dynamical time, $t_{\text dyn}^{\text mj}$, of the micro-jet and the moment when it was launched. If the projected distance of the micro-jet from the star is $l\approx 1\arcsec \approx 130$~au \citep{Nisini-2023} and the jet tangential velocity relative to BP~Tau is $V_{\text t}^{\text mj}= V_{\text r}^{\text mj} \tan i \approx 86$~\kms, where $i=38\fdg1\pm 0\fdg5$ is the inclination of the disk axis to the line of sight \citep{Long-2019}, then $t_{\text dyn}^{\text mj}=l/V_{\text t}^{\text mj} \approx 7.2$~yr. 

Therefore, the micro-jet was launched in the second half of 2016 and moved along the minor axis of the ALMA disk image with a projected velocity of $V_{\text t}\approx 0\farcs14$ yr$^{-1}.$ For a certain period of time, $\Delta t$, the counter-jet will be eclipsed by the disk, that is, not visible. We cannot estimate $\Delta t$ because the radius of the disk region, $R_{\text op},$ which is opaque to radiation at $\lambda \approx 672$~nm, is not known. If $R_{\text op}$ is equal to the spatial extent of the semiminor axis, $R_{\text A}$, of the ALMA image at $\lambda=1.3$~mm (0\farcs25 according to \citealt{Long-2019}), then $\Delta t\approx 2$~yr. However, it is not clear how different the values of $R_{\text op}$ and $R_{\text A}$ are. 

About 3.6 years after the launch, that is, since about the middle of 2020, the jet has reached a distance of more than 0\farcs5 and thus left the slit of spectrographs with a smaller aperture, for example ESPRESSO or HARPS-N, used by \citet{Nisini-2023}. We believe this to be the reason for the absence of [\ion{S}{ii}] lines in the high-resolution spectra observed in 2021-2022. 

\begin{figure}
   \centering
   \includegraphics[width=\hsize]{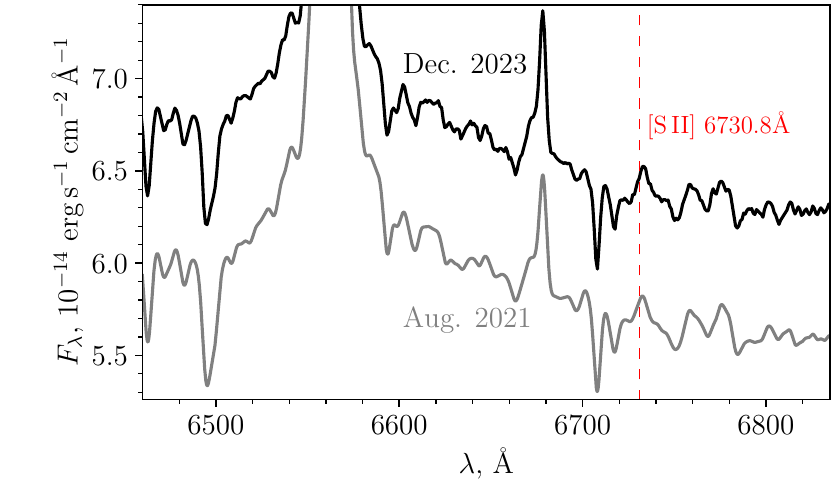}
      \caption{Comparison of BP~Tau spectra observed in August 2021 (X-shooter, 5\arcsec\  slit, gray curve) and in December 2023 (TDS, 10\arcsec\ slit).}
         \label{fig:xshoo}
\end{figure}

\begin{figure*}
   \centering
   \includegraphics[width=\hsize]{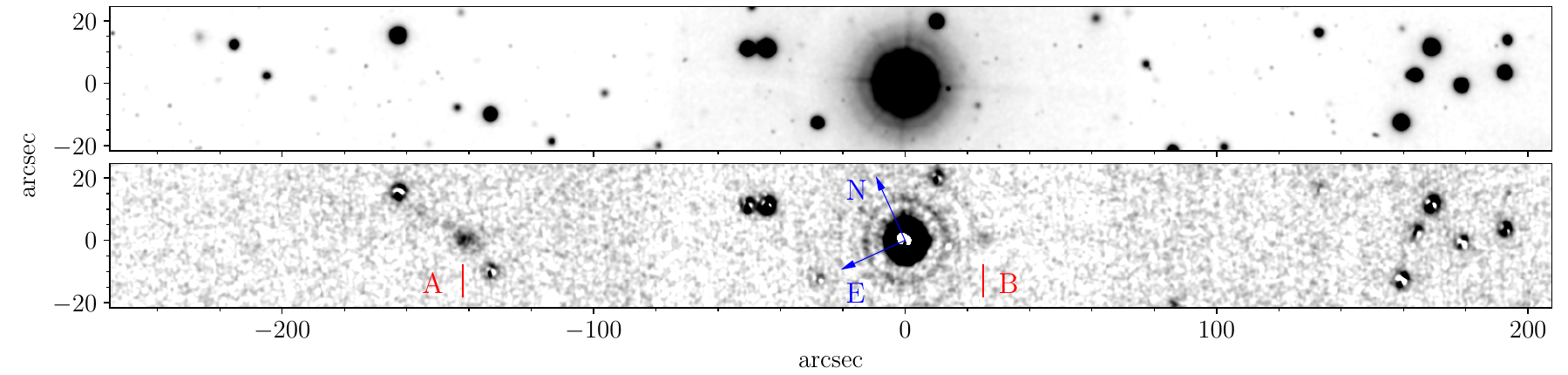}
      \caption{Vicinity of BP~Tau along the direction of PA$=59\degr$ in the continuum band [\ion{S}{ii}]rc (upper panel) and the difference between images in the [\ion{S}{ii}] and [\ion{S}{ii}]rc filters (lower panel). Red markers indicate the location of the HH objects (knots A and B).}
         \label{fig:jet_img}
\end{figure*}

On the other hand, if $R_{\text op}<0\farcs5,$ then the lines should have been present in wider-slit spectra in the second half of 2020, and we did find an X-shooter spectrum in the ESO archive observed in August 2021 in which the [\ion{S}{ii}]~6731~\AA{} line is present; it has the same flux, $(6\pm1)\times 10^{-15}$~erg\,s$^{-1}$\,cm$^{-2}$, as in the TDS spectra from December 2023 (see Fig.\,\ref{fig:xshoo}, where we have reduced the resolution of the X-shooter spectrum to that of the TDS one). Both spectra were observed with wide slits, and therefore the micro-jet was completely covered by the apertures.

\subsection{The jet}
 \label{subsec:jet}   
A part of the $10\arcmin \times 10\arcmin$ [\ion{S}{ii}]rc image of the BP~Tau vicinity is shown in the upper panel of Fig.~\ref{fig:jet_img}. The difference between the images of the same region in the [\ion{S}{ii}] and [\ion{S}{ii}]rc filters (see Sect.\,\ref{sec:obs}) is shown in the bottom panel of the figure. One can see two faint diffuse knots (A and B) located on either side of the star with a S/N per square-arcsecond of $\approx 5$  and $\approx 2$ for knots A and B, respectively. The characteristic size of knot A is $\approx 5\arcsec.$ The absolute flux of the brightest part of knot A within the circular aperture with a diameter of $5\arcsec$ is $8\times 10^{-16}$~erg~s$^{-1}$ cm$^{-2}$~\AA$^{-1}$ for the [\ion{S}{ii}] lines; knot B is about three times fainter in the same aperture.

\begin{figure}
   \centering
   \includegraphics[width=\hsize]{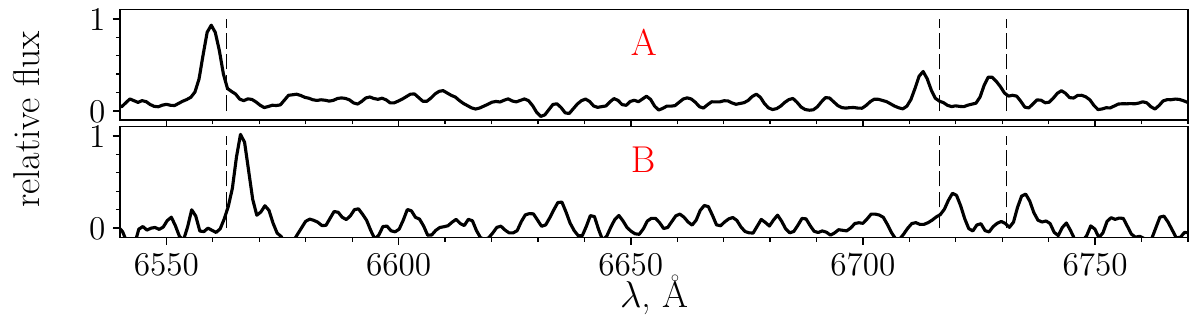}
      \caption{TDS spectra of the knots (see Table\,\ref{tab:log-TDS}; 1\farcs5 slit). Dashed lines mark the rest positions of the H$\alpha$ and [\ion{S}{ii}] lines. They indicate that the emission lines have the same blueshift and redshift (see Table\,\ref{tab:jet-knots}).}
         \label{fig:jet_sp}
\end{figure}

The red-channel TDS spectra of these knots (Fig.~\ref{fig:jet_sp}) indicate that the knots are emission nebulae radiating in the H$\alpha,$ [\ion{S}{ii}]~6716, and 6731~\AA{} lines.
The observed H$\alpha$ to [\ion{S}{ii}] flux ratio in the knots is nearly 1.5, which is typical for HH objects (see \citealt{Dopita-2017} and references therein).
These lines are blueshifted for knot A and redshifted for knot B. Equatorial coordinates, the distances from BP~Tau, and radial velocities (with the $V_{\text r}$ of the star subtracted) are presented in Table~\ref{tab:jet-knots}. 

The PA of the line connecting knot B, BP~Tau, and knot A is $59 \pm 1\degr,$ which practically coincides with the PA=$61 \pm 1\degr$ of the minor axis of the ALMA disk image \citep{Long-2019} and coincides within measurement errors with the micro-jet PA. Furthermore, the BP~Tau radial velocity of $16.09 \pm 0.16$~{\kms} \citep{Nisini-2023} is much lower than that of the knots, which means that the knots move from the star in opposite directions. On these grounds, we believe the knots to be HH objects in the jet (knot A) and counter-jet (knot B) from BP~Tau. Professor Bo~Reipurth agreed with our arguments; he included the HH flow from BP~Tau in his catalog \citep{Reipurth-2000} and named it HH~1181.

One can estimate the mass-loss rate of the micro-jet as follows \citep{Dopita-2017}. According to Sect.~\ref{subsec:mjet}, $N_{\text e}$ is $\gtrsim 10^4$~cm$^{-3}$ in the [\ion{S}{ii}] line formation region. The typical transverse radius of micro-jets, $r_{\text mj}$, is $\gtrsim 10$~au, and the hydrogen ionization degree, $x_{\text e}=N_{\text e}/N_{\text H}$, is nearly 0.1 in the [\ion{S}{ii}] line formation region \citep{Cabrit-2007}. Thus, 
\begin{equation}
\dot M_{\text mj} \approx  \pi r_{\text mj}^2 m_{\rm H} \frac{N_{\text e}}{x_{\text e}} V_{\text mj}
\gtrsim 2 \times 10^{-9} \,\, \mbox{M}_\odot \,\, \mbox{yr}^{-1},
\label{eq:Mdot}
\end{equation}
where $V_{\text mj}=V_{\text r}^{\text j}/\cos i \approx 140$~km\,s$^{-1}$ is the velocity of the micro-jet relative to the star and $m_{\rm H}$ is the mass of a hydrogen atom. According to \citet{Nisini-2023}, the accretion rate for BP~Tau is $\dot M_{\text ac}=6\times 10^{-9}$~M$_\odot$~yr$^{-1},$ that is, its  $\dot M_{\text mj}/  \dot M_{\text ac}$ ratio is $\approx 0.3,$ which is close to the typical value of $\sim 0.1$ \citep{Frank-2014}, taking the possible uncertainty of the adopted parameters into account. Unfortunately, we cannot estimate $\dot M$ for knots A and B  due to a very uncertain upper limit for $N_\text{e}.$ 

As follows from Fig.\,\ref{fig:jet_sp}, the flux ratio, $f_{\text S}$, for the [\ion{S}{ii}] 6716 and 6731~\AA{} lines is close to unity, whereas in the case of the micro-jet, $f_{\text S}$ is noticeably less than unity. We do not have enough data to determine the physical conditions in the studied objects, but judging by these $f_{\text S}$ values, the electron number density, $N_{\text e}$, in the HH A and B objects (knots) is noticeably lower than in the micro-jet (see, e.g., \citealt{Giannini-2015}).

\begin{table}
\caption{Parameters of HH objects.}
\label{tab:jet-knots}                     
\centering                          
\begin{tabular}{c c c c c c }   
\hline\hline                        
Object & $\alpha_{2000}$ & $\delta_{2000}$ & $d,$ \arcsec & $V_{\text r},$ \kms \\
\hline                       
Knot A & 04:19:25.05 & 29:07:39.7 & 141.1 & $-160 \pm 5$  \\ 
Knot B & 04:19:14.13 & 29:06:14.7 & 25.4  & $+138 \pm 7$  \\ 
\hline                                   
\multicolumn{6}{l}{$d$ is the distance from the star. $V_{\text r}$ is the radial velocity;}\\ 
\multicolumn{6}{l}{the stellar velocity, +16\,{\kms}, has been removed.}\\
\end{tabular}\\
\end{table}

As was noted in the Introduction, the [\ion{S}{ii}]~6731~\AA{} line, redshifted by $\approx 120$~{\kms} relative to the stellar rest frame, was observed in the BP~Tau spectrum by \citet{HEG-1995} in January 1988 (i.e., 36 years ago). It follows from Table\,\ref{tab:jet-knots} that the age of knots A and B is $t_{\text dyn}^{\text A}\approx 700$ and $t_{\text dyn}^{\text B} \approx 145$ years, respectively. Judging by the dynamical time, $t_{\text dyn}^{\text B}$, in 1988 knot B was at a distance of $\approx 19\arcsec$ from the star, it could not have fallen into the spectrograph slit $(1\farcs25 \times 7\farcs0$).
It is likely that one more HH object exists in the counter-jet at a current distance of 5-10{\arcsec} from the star, but we have not detected it in the direct image shown in Fig.~\ref{fig:jet_img} (perhaps due to a bright halo around BP~Tau) or in the TDS spectra. Additional observations with better angular resolution and sensitivity are needed to test this hypothesis.


\section{Conclusion}
 \label{sec:conclusion}

We have discovered the HH outflow from BP~Tau, which was assigned the number HH~1181 in the \citet{Reipurth-2000} catalog. Based on our observations, we have come to the following conclusions.
   \begin{enumerate}
      \item The outflow is highly collimated and consists of two HH objects and a micro- (counter-) jet.
      \item The outflow is directed practically along the rotation axis of the BP~Tau protoplanetary disk.  
      \item The jet from BP~Tau is bipolar, but asymmetric: the number of HH objects as well as their distances and velocities relative to the star are different in oppositely directed parts of the outflow. 
      Most jets from CTTSs are asymmetric, so BP~Tau is apparently not an exception. What we see as a jet in fact consists of a string of HH objects; the initial asymmetry possibly originates in the  {\lq\lq}central engine{\rq\rq} and evolves into the asymmetric appearance of HH objects.
      \item  There may be one more HH object in the counter-jet of BP~Tau at a distance of $5-10\arcsec$ from the star.
   \end{enumerate}
   

\begin{acknowledgements}
We thank the staff of the CMO SAI MSU for the help with observations and Prof. Bo~Reipurth for including the discovered HH-flow in the general catalog of objects of this type and assigning it the number HH~1181. We also thank the referee for useful remarks. This research has made use of the SIMBAD database (CDS, Strasbourg, France), Astrophysics Data System (NASA, USA), NIST Atomic Spectra Database (https://www.nist.gov/pml/atomic-spectra-database) and The atomic line list v2.04 (https://linelist.pa.uky.edu/atomic/). This work was conducted under the financial support from the Russian Science Foundation (grant 23-12-00092). Scientific equipment used in this study was bought partially through the M.~V.~Lomonosov Moscow State University Program of Development.
\end{acknowledgements}


\bibliographystyle{aa}  
\bibliography{bptau}    

\end{document}